\documentclass{PoS}
\usepackage{subcaption}
\usepackage{graphicx}
\usepackage{mathtools}

\title{Higgs boson as a gluon trigger: the study of QCD in high pile-up environments}

\ShortTitle{Higgs boson as a gluon trigger}

\author{\speaker{H. Van Haevermaet}$^{,a,b}$, P. Cipriano$^{a}$, S. Dooling$^{a}$, A. Grebenyuk$^{c}$, P. Gunnellini$^{a,b}$, F.~Hautmann$^{d,e,f}$, H. Jung$^{a,b}$, P. Katsas$^{a}$\\
\llap{$^{a}$}Deutsches Elektronen Synchrotron, D 22603 Hamburg\\
\llap{$^{b}$}Universiteit Antwerpen, Elementaire Deeltjes Fysica, B 2020 Antwerpen\\
\llap{$^{c}$}Universit\'e Libre De Bruxelles, Institut Interuniversitaire des Hautes Energies, B 1050 Bruxelles\\
\llap{$^{d}$}University of Sussex, Physics \& Astronomy, Brighton BN1 9QH\\
\llap{$^{e}$}University of Oxford, Physics Department, Oxford OX1 3NP\\
\llap{$^{f}$}Rutherford Appleton Laboratory, Chilton OX11 0QX\\
E-mail: \email{hans.vanhaevermaet@uantwerpen.be}}

\abstract{In the forthcoming high-luminosity phase of the LHC many of the most interesting measurements for precision QCD studies are hampered by large pile-up conditions, especially at not very high transverse momenta. However, with the recently discovered Higgs boson, which couples in the heavy top limit directly to gluons, we have access to a novel production process to probe QCD by a colour-singlet current. In this study we compare observables in Higgs boson and Drell-Yan production and investigate whether measuring ratios or subtractions can yield results that are stable in high pile-up environments, and yet sensitive to (small-$p_{\text{T}}$) QCD physics in gluon fusion processes. We present results of Monte Carlo event generator calculations for a few specific examples.}

\FullConference{XXII. International Workshop on Deep-Inelastic Scattering and Related Subjects \\
		 28 April - 2 May 2014 \\
		 Warsaw, Poland}

\begin{document}

\section{Introduction}
To study QCD one typically measures processes in single proton-proton collisions. A particular interesting one is the Drell-Yan $q\bar{q} \xrightarrow{\gamma^{*}/Z^{0}} l^{+}l^{-} $ process, which is produced through an electroweak current that couples to quarks. A big advantage is its clean final state, which only involves the decay leptons. This allows us to use the process to precisely measure the quark structure functions, quark induced parton showers, and underlying event properties. The properties and structure functions of gluons however, have to be determined indirectly. But with the recent discovery \cite{bib:higgs} of the Higgs boson, this picture changes. In the heavy top limit, the Higgs boson directly couples to gluons and can be produced with the gluon fusion $gg \rightarrow H$ process through a colour singlet current \cite{bib:gluonfusion}. The access to this new production process opens up a whole new interesting area of possible QCD measurements, where we can directly study gluon induced effects. In addition, if we only look at decay channels of the Higgs boson like $H \rightarrow ZZ \rightarrow 4l$, we have access to the same clean final state as in the Drell-Yan process. In order to produce enough statistics for detailed measurements, current accelerators like the LHC at CERN have to operate at very high beam intensities. This high luminosity allows one to measure differential distributions, e.g. the transverse momentum \cite{bib:atlasnote} of the Higgs, but it also creates a condition in which on average much more than one proton-proton collision happens per bunch crossing. This so called pile-up (PU) can reach, given the LHC run conditions, scenarios where PU = 20 (i.e. 20 additional proton-proton collisions occur on top of the one signal event). This implies that the phase space will be completely filled with extra hadronic activity coming from the pile-up events, which makes a study of QCD in the Higgs or Drell-Yan channels extremely difficult. To assess whether one can still perform QCD measurements in such harsh environments we initiated a new program using the Higgs boson as a gluon trigger \cite{bib:higgsgluon}. The main idea is to compare Higgs and Drell-Yan production in the same invariant mass range, and then look at different observables, such as the transverse momentum of the bosons, to measure the difference in soft multi-gluon emissions. In these proceedings we present pile-up studies to show whether the Higgs to Drell-Yan comparison stays valid in high pile-up environments. We look at the ratio (Higgs/DY) of observables, as it is sensitive to the direct difference in soft gluon versus quark radiation, and we look at the subtraction (Higgs - DY) to determine if one can remove the PU contributions from the underlying event.

The Monte Carlo event generator samples used in this study are all produced with Pythia 8.185 \cite{bib:pythia8}. All samples contain proton-proton collisions produced at a centre-of-mass energy of $\sqrt{s} = 7$~TeV with the 4C tune \cite{bib:4Ctune} to describe the underlying event properties. The Higgs production sample is generated by activating the gluon fusion process $gg\rightarrow H$, while the Drell-Yan sample is generated by using the single $Z^{*}$ production process $f\bar{f} \rightarrow Z^{0*}$. The bosons are produced within the same invariant mass range of $115 < \text{M} < 135$~GeV and in addition, to avoid complications with the leptonic final states, they are set stable. The Higgs mass is set to 125~GeV. To study the effect of additional pile-up events, samples with a fixed amount of PU = 5 and PU = 20 are produced. This is done by adding a specified number of small-$p_{\text{T}}$ QCD process events to the signal event. All jets in the samples are constructed with the anti-$k_{\text{T}}$ algorithm \cite{bib:antikt} with a distance parameter R = 0.5.

\section{Transverse momentum spectra}
\begin{figure}[h]
        \centering
        \begin{subfigure}[b]{0.50\textwidth}
                \centering
                \includegraphics[width=\textwidth]{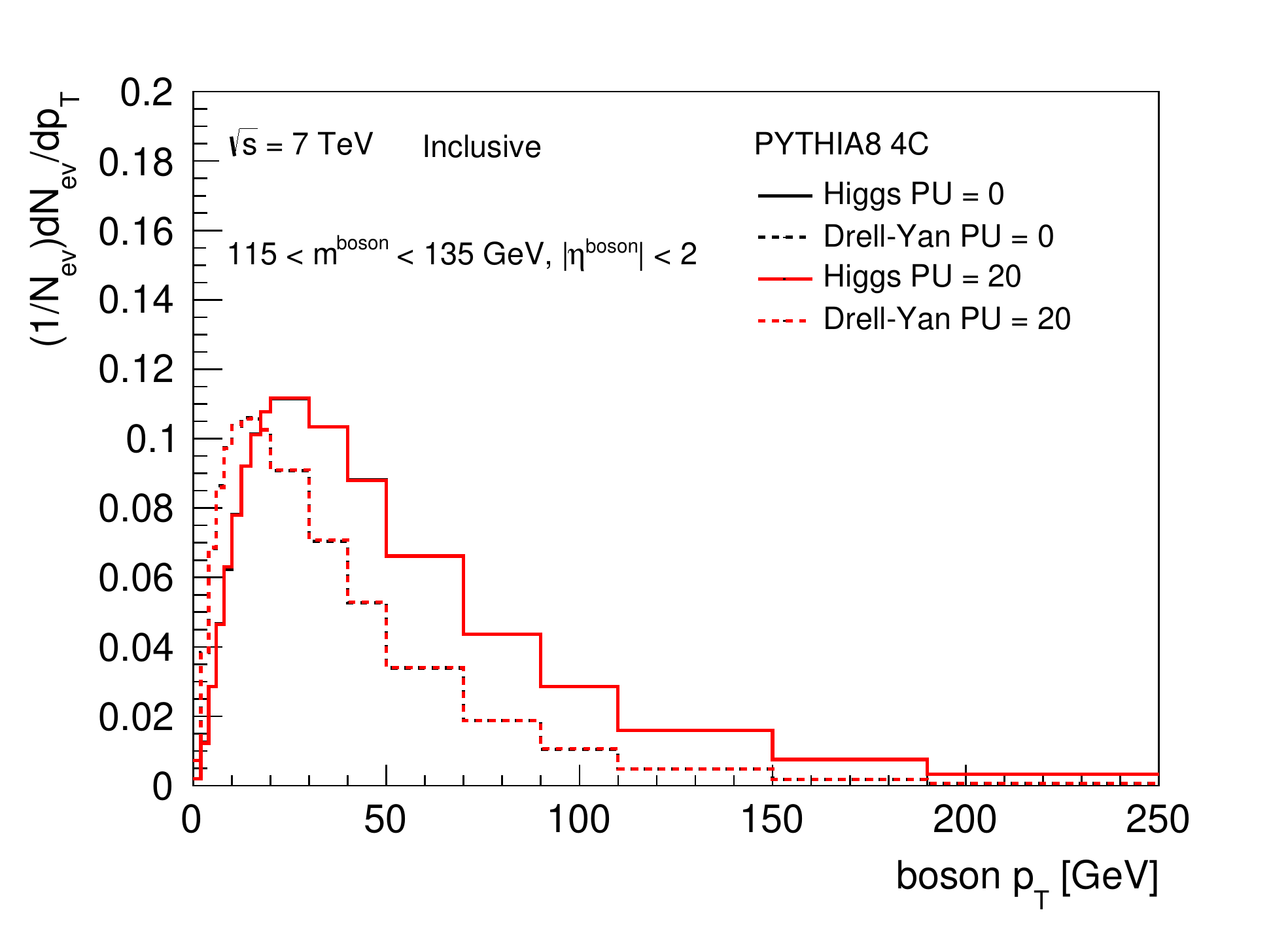}
                \caption{}
                \label{fig:DefCompPU_inclusive}
        \end{subfigure}%
        ~ 
        \begin{subfigure}[b]{0.50\textwidth}
                \centering
                \includegraphics[width=\textwidth]{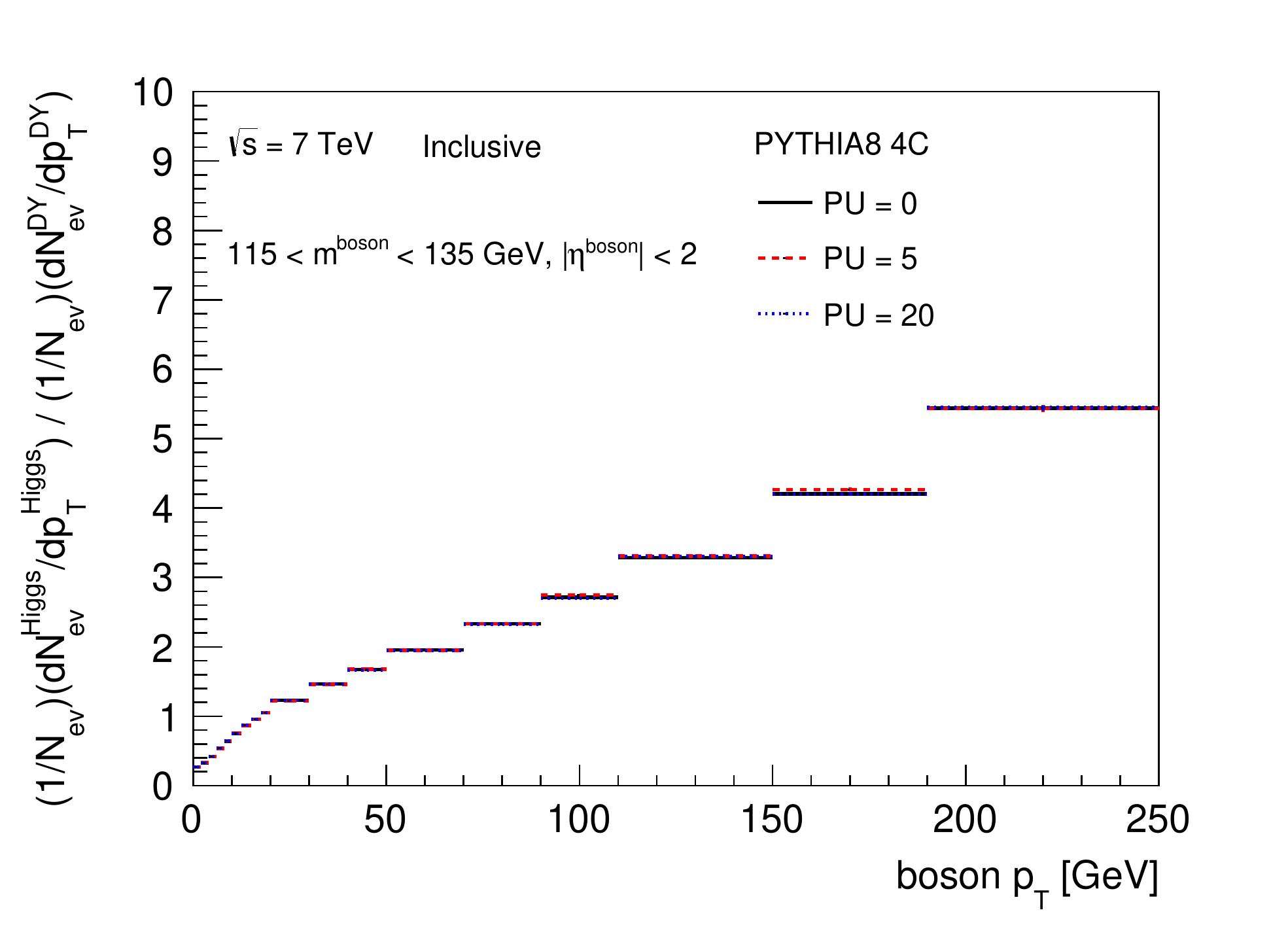}
                \caption{}
                \label{fig:Ratio_inclusive}
        \end{subfigure}
        \caption{(a) The inclusive Higgs (solid lines) and Drell-Yan (dashed lines) transverse momentum $p_{\text{T}}$ spectra for PU = 0 (black lines) and PU = 20 (red lines) scenarios. (b) The ratio of the inclusive Higgs and Drell-Yan transverse momentum $p_{\text{T}}$ spectra for PU = 0 (solid black), PU = 5 (dashed red) and PU = 20 (dotted blue) scenarios.}
        \label{fig:PTinclusive}
\end{figure}

Initial studies presented in \cite{bib:higgsgluon} show the implicit difference in the inclusive $p_{\text{T}}$ spectra of Higgs and Drell-Yan production, due to the different soft gluon emissions. Here we analyse what happens when one includes additional PU events. By definition, as we put the bosons stable, the $p_{\text{T}}$ spectra should be independent. This is confirmed in figure \ref{fig:DefCompPU_inclusive}, where both the PU = 0 and PU = 20 scenarios are compared for Higgs and Drell-Yan production within $\left|\eta\right| < 2$ and $115 < \text{M} < 135$~GeV. Thus, if the experimental reconstruction of the leptonic decay products is stable in high PU environments, the $p_{\text{T}}$ spectrum should be stable. One can then take the ratio (Higgs/DY) of both spectra and use this observable to directly quantify the gluon versus quark radiation effects, which generate the transverse momenta of the bosons. This is shown in figure \ref{fig:Ratio_inclusive} for the PU = 0, PU = 5 and PU = 20 scenarios and is indeed found to be stable in high PU environments. 

\begin{figure}[h]
        \centering
        \begin{subfigure}[b]{0.50\textwidth}
                \centering
                \includegraphics[width=\textwidth]{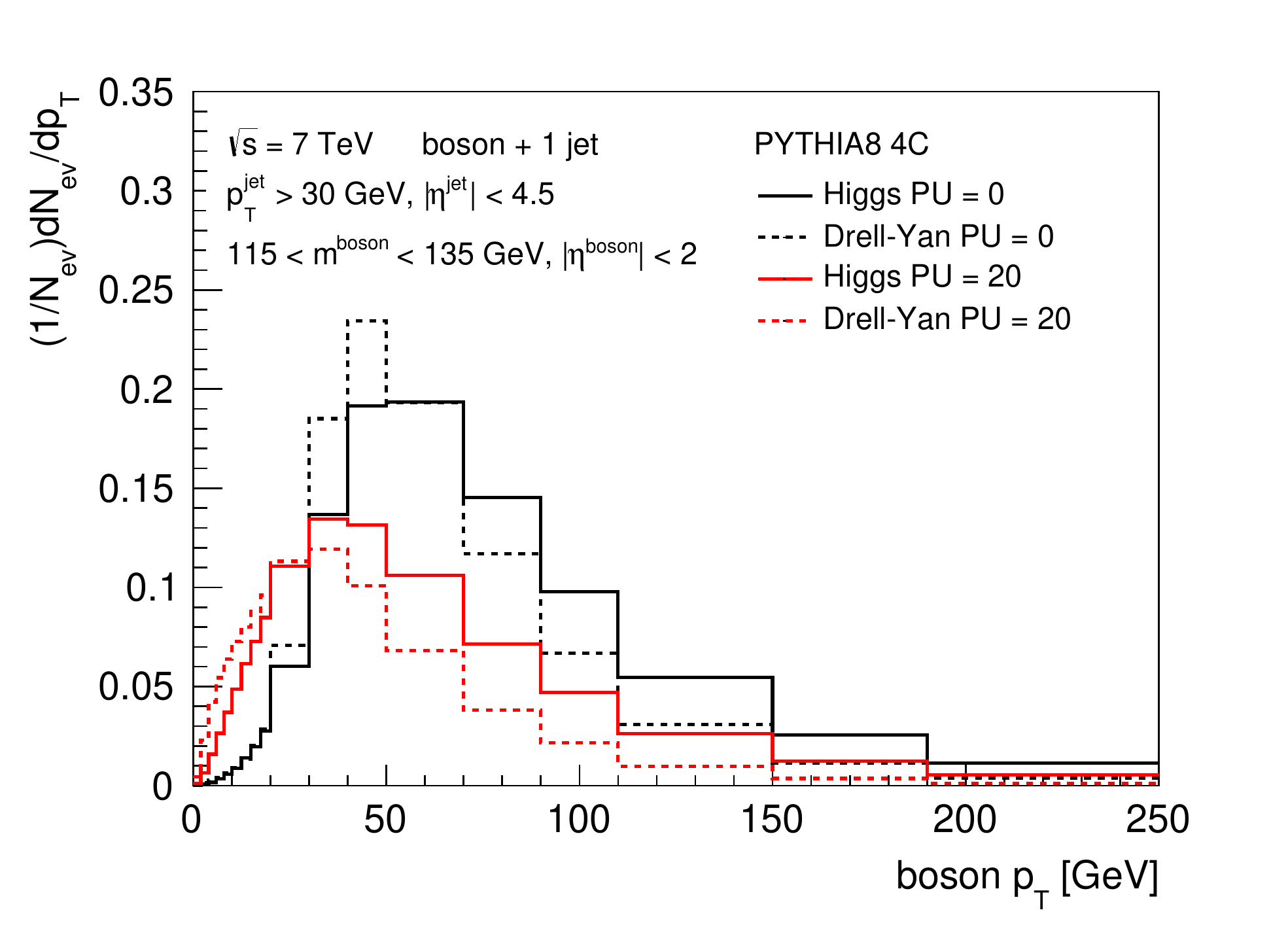}
                \caption{}
                \label{fig:DefCompPU_1jet}
        \end{subfigure}%
        ~ 
        \begin{subfigure}[b]{0.50\textwidth}
                \centering
                \includegraphics[width=\textwidth]{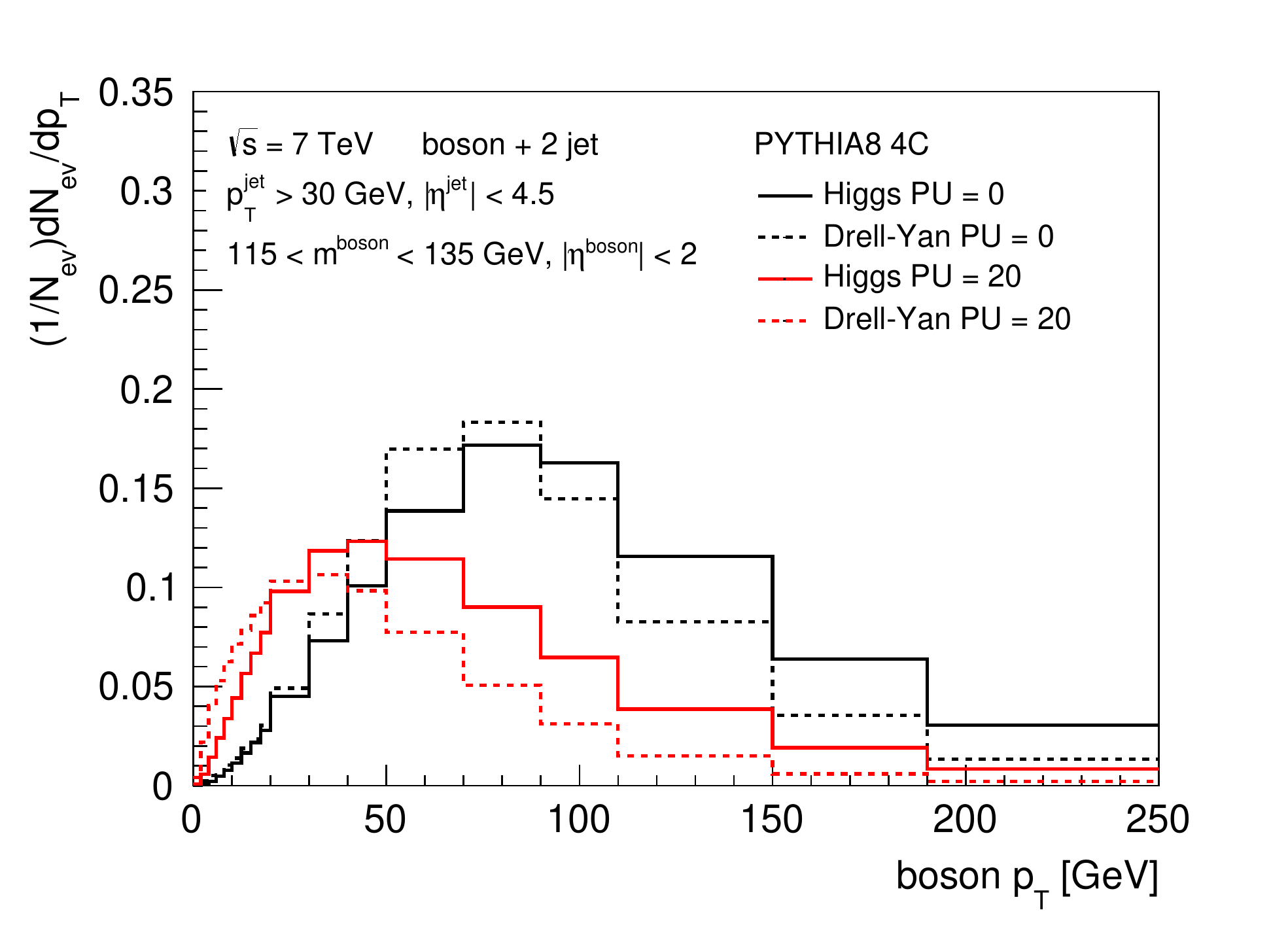}
                \caption{}
                \label{fig:DefCompPU_2jet}
        \end{subfigure}
        \caption{(a) The Higgs (solid lines) and Drell-Yan (dashed lines) + 1 jet production transverse momentum $p_{\text{T}}$ spectra for PU = 0 (black lines) and PU = 20 (red lines) scenarios. (b) The Higgs (solid lines) and Drell-Yan (dashed lines) + 2 jet production transverse momentum $p_{\text{T}}$ spectra for PU = 0 (black lines) and PU = 20 (red lines) scenarios.}
        \label{fig:PTPUjets}
\end{figure}

In addition one can look at the following event topologies: boson + 1 jet and boson + 2 jet production. These topologies are expected to be even more sensitive to gluon versus quark emission effects, as hard radiation accompanies the boson production. However requiring additional hard jets ($p_{\text{T}} > 30$~GeV/c, $\left|\eta\right| < 4.5$) shifts the spectra towards higher transverse momenta as a result of the induced $p_{\text{T}}$ balance between the boson and the hard jets. As such, the contribution of the gluon versus quark induced effects will actually become less pronounced in the $p_{\text{T}}$ spectra. These effects are illustrated in figures \ref{fig:DefCompPU_1jet} and \ref{fig:DefCompPU_2jet} that show the boson + 1 jet and boson + 2 jet topologies respectively. Comparing the aforementioned figures with figure \ref{fig:DefCompPU_inclusive} it is clear that the average $p_{\text{T}}$ increases when requiring additional hard jets, and that the difference between the Higgs and Drell-Yan spectra becomes smaller. The latter can be quantified by calculating the ratio $\left<p_{\text{T}}\right>^{\text{Higgs}}/\left<p_{\text{T}}\right>^{\text{DY}}$ for the inclusive, boson + 1 jet and boson + 2 jet scenarios, which is 1.52, 1.17 and 1.16 respectively. Furthermore, figures \ref{fig:DefCompPU_1jet} and \ref{fig:DefCompPU_2jet} also show the results of adding additional PU events. In this case, both the Higgs and Drell-Yan spectra shift to lower values. This is the consequence of the jet mismatching that occurs due to the extra PU activity. That is, in the presence of PU, there is a higher probability that a high $p_{\text{T}}$ jet comes from an independent PU interaction and is identified as the jet coming from the signal event, which can result in a matching of a high $p_{\text{T}}$ jet with a low $p_{\text{T}}$ boson.

\begin{figure}[h]
        \centering
        \begin{subfigure}[b]{0.50\textwidth}
                \centering
                \includegraphics[width=\textwidth]{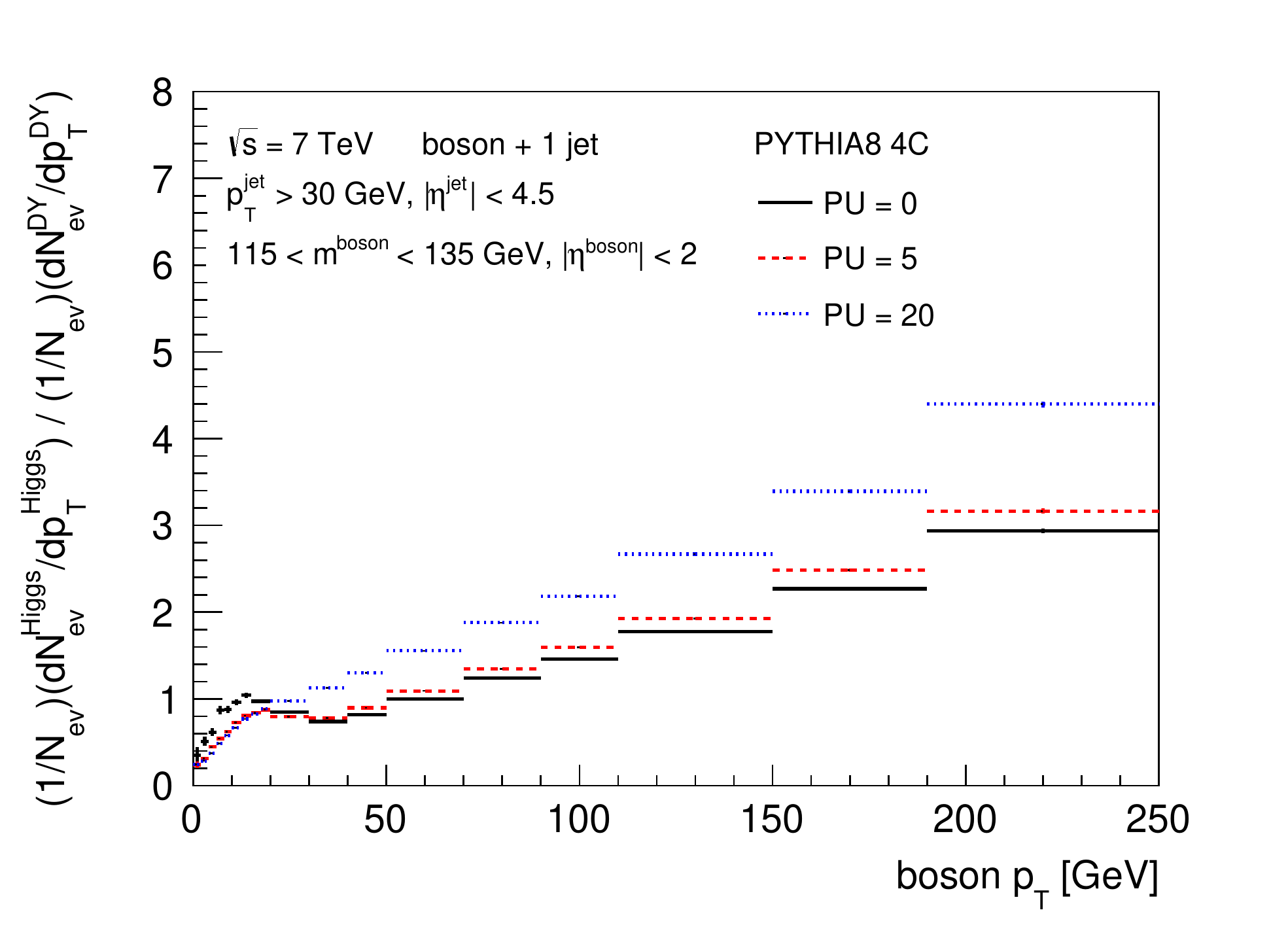}
                \caption{}
                \label{fig:Ratio_1jet}
        \end{subfigure}%
        ~ 
        \begin{subfigure}[b]{0.50\textwidth}
                \centering
                \includegraphics[width=\textwidth]{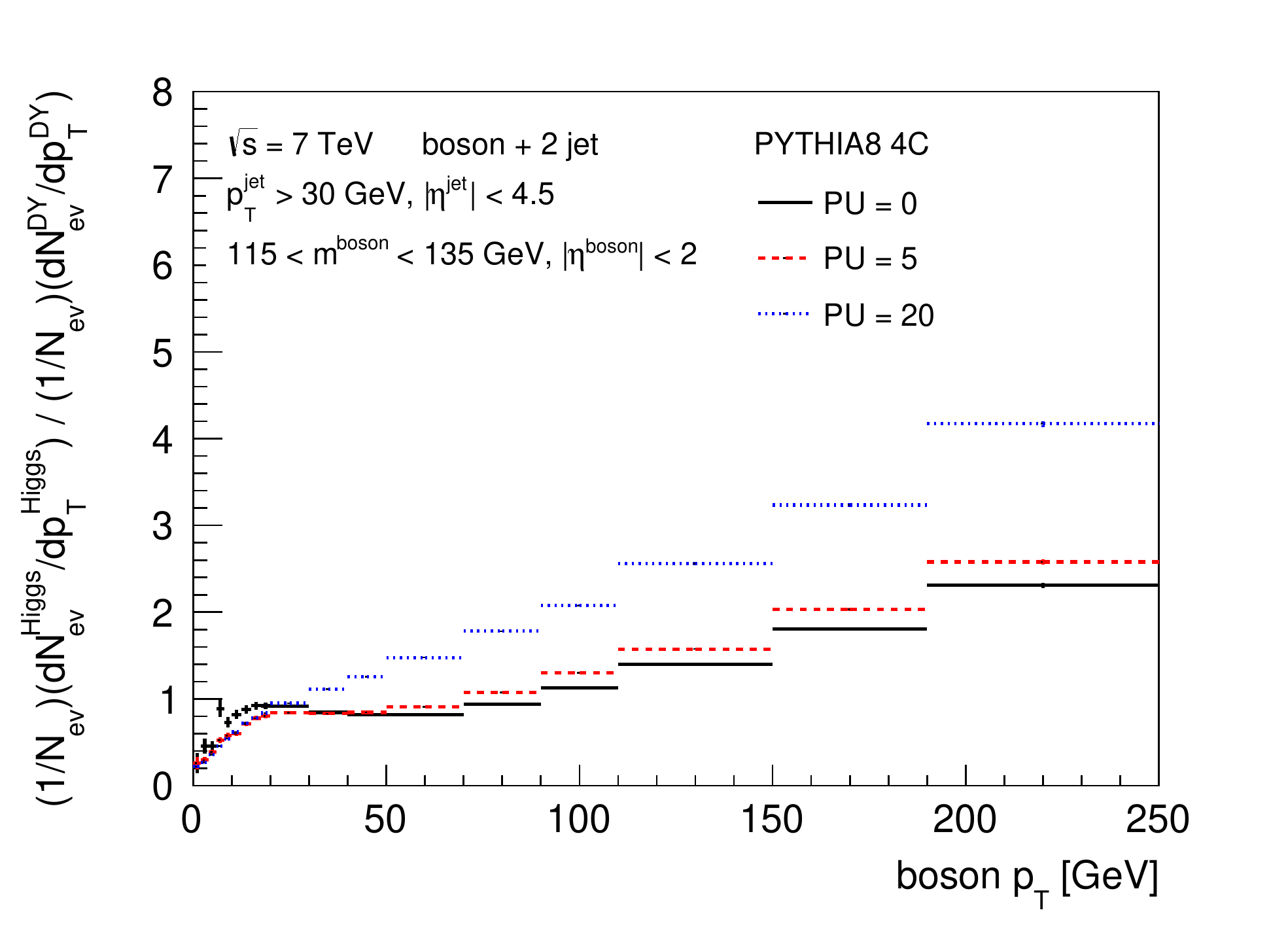}
                \caption{}
                \label{fig:Ratio_2jet}
        \end{subfigure}
        \caption{(a) The ratio of Higgs and Drell-Yan + 1 jet production transverse momentum $p_{\text{T}}$ spectra for PU = 0 (solid black), PU = 5 (dashed red) and PU = 20 (dotted blue) scenarios. (b) The ratio of Higgs and Drell-Yan + 2 jet production transverse momentum $p_{\text{T}}$ spectra for PU = 0 (solid black), PU = 5 (dashed red) and PU = 20 (dotted blue) scenarios.}
        \label{fig:PTRatios}
\end{figure}

As with the inclusive spectrum one can check whether the ratio (Higgs/DY) of the spectra is stable with additional PU activity. This is shown in figures \ref{fig:Ratio_1jet} and \ref{fig:Ratio_2jet}. However, due to the different inclusive $p_{\text{T}}$ spectrum between Higgs and Drell-Yan production, the fraction of leading jet mismatching is not the same, and hence the ratios of the boson + 1 jet and boson + 2 jet $p_{\text{T}}$ spectra are not stable with respect to PU effects. 

\section{Underlying event observables}
\begin{figure}[h]
        \centering
        \begin{subfigure}[b]{0.50\textwidth}
                \centering
                \includegraphics[width=\textwidth]{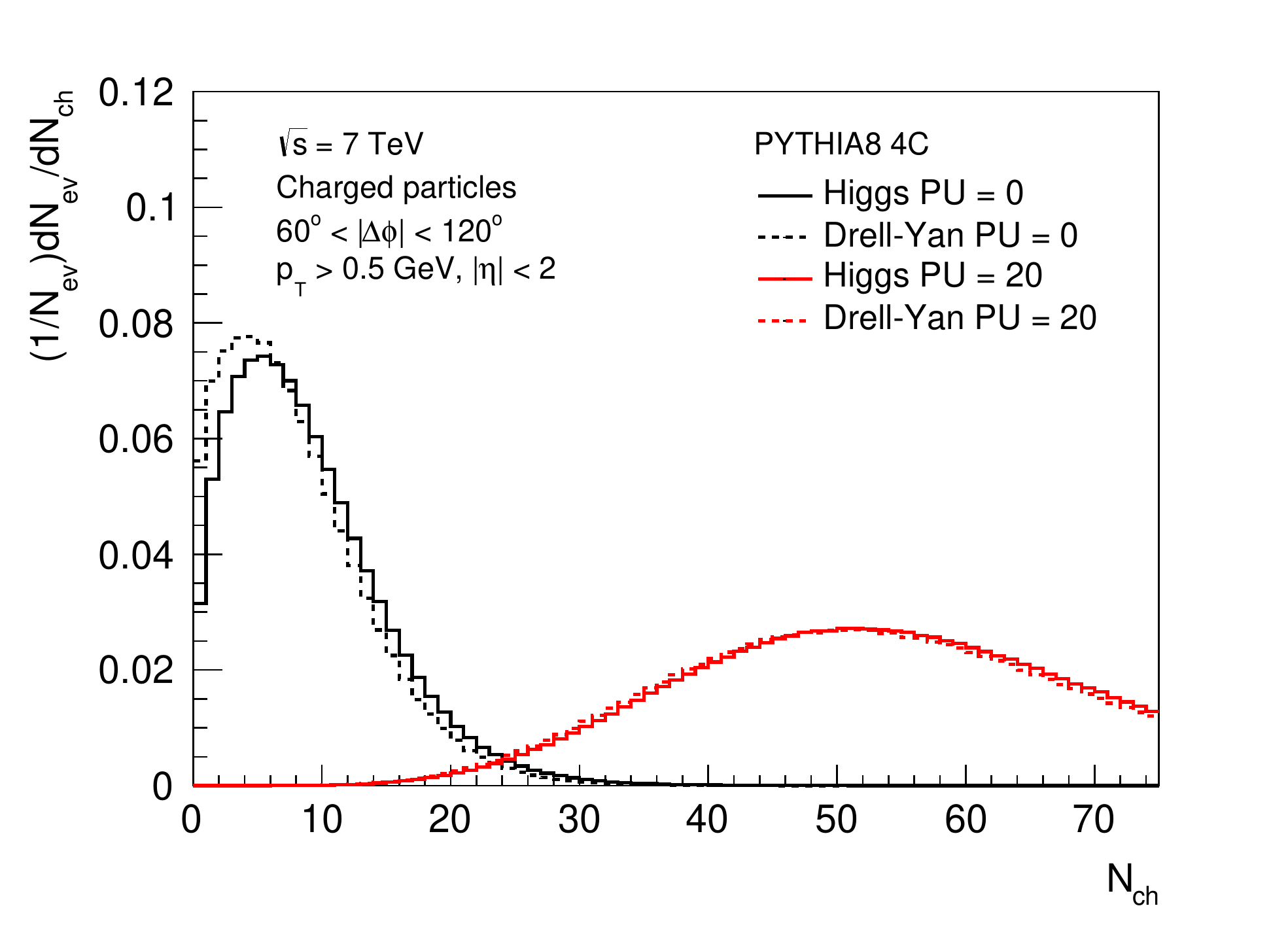}
                \caption{}
                \label{fig:DefCompPU_ChPartmulti}
        \end{subfigure}%
        ~ 
        \begin{subfigure}[b]{0.50\textwidth}
                \centering
                \includegraphics[width=\textwidth]{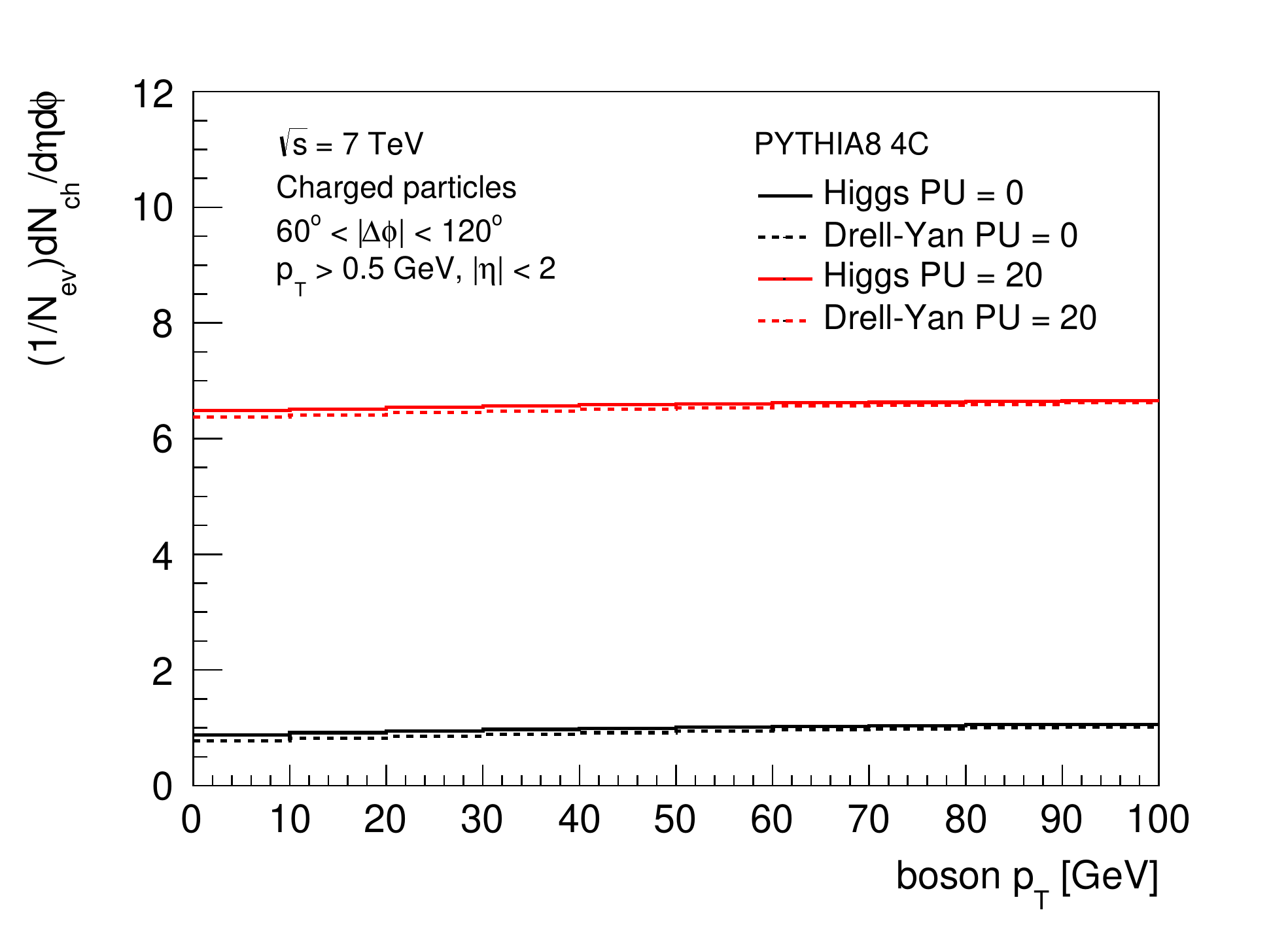}
                \caption{}
                \label{fig:DefCompPU_ChPartmulti_vs_Bpt}
        \end{subfigure}
        \caption{(a) The charged particle multiplicity in the transverse region of the azimuthal plane for Higgs (solid lines) and Drell-Yan (dashed lines) production for PU = 0 (black lines) and PU = 20 (red lines) scenarios. (b) The average charged particle multiplicity in the transverse region of the azimuthal plane versus Higgs (solid lines) and Drell-Yan (dashed lines) transverse momentum $p_{\text{T}}$ for PU = 0 (black lines) and PU = 20 (red lines) scenarios.}
        \label{fig:UE}
\end{figure}

\begin{figure}[h]
        \centering
        \begin{subfigure}[b]{0.50\textwidth}
                \centering
                \includegraphics[width=\textwidth]{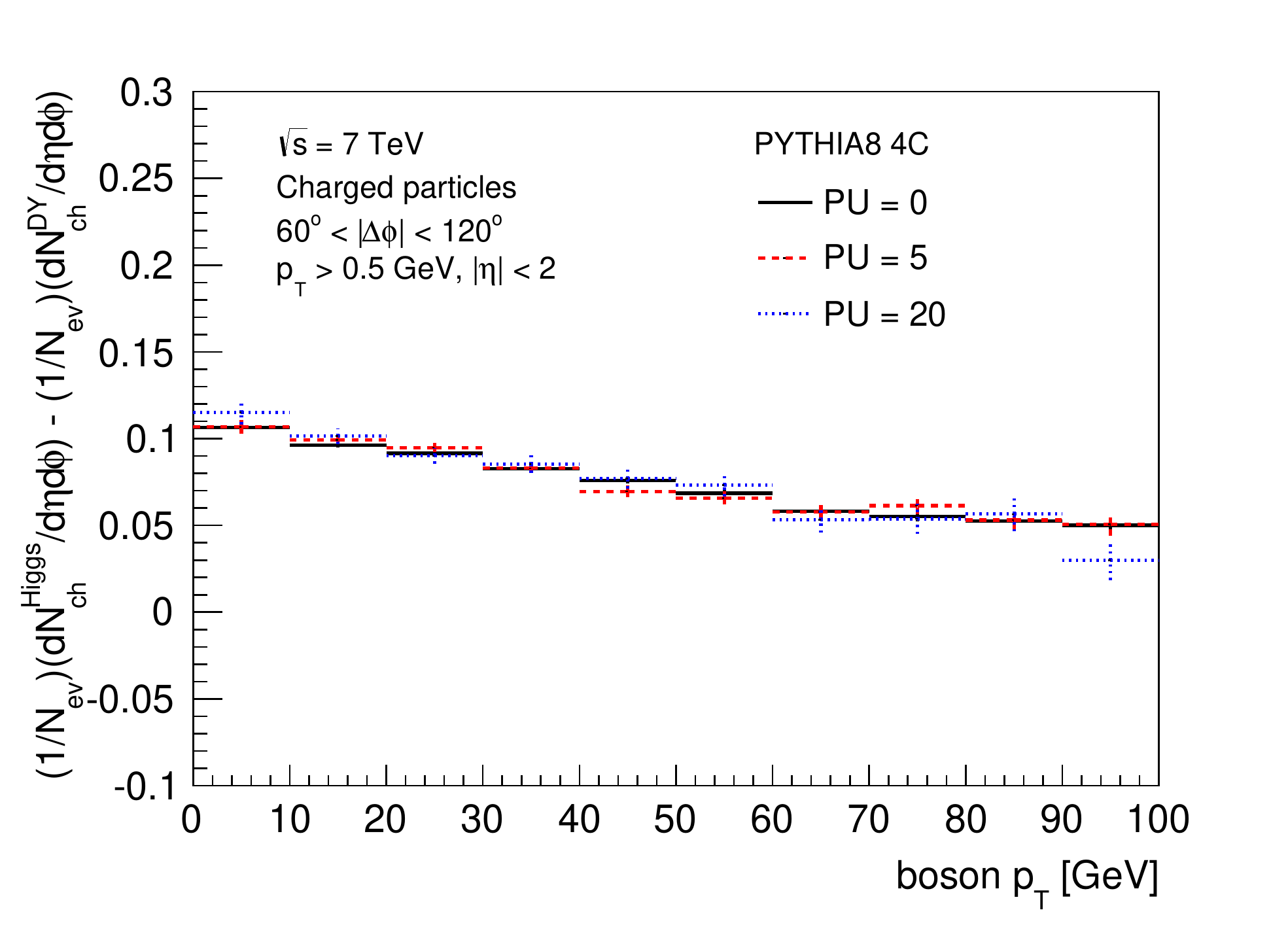}
                \caption{}
                \label{fig:Sub_ChPartmulti_vs_Bpt}
        \end{subfigure}%
        ~ 
        \begin{subfigure}[b]{0.50\textwidth}
                \centering
                \includegraphics[width=\textwidth]{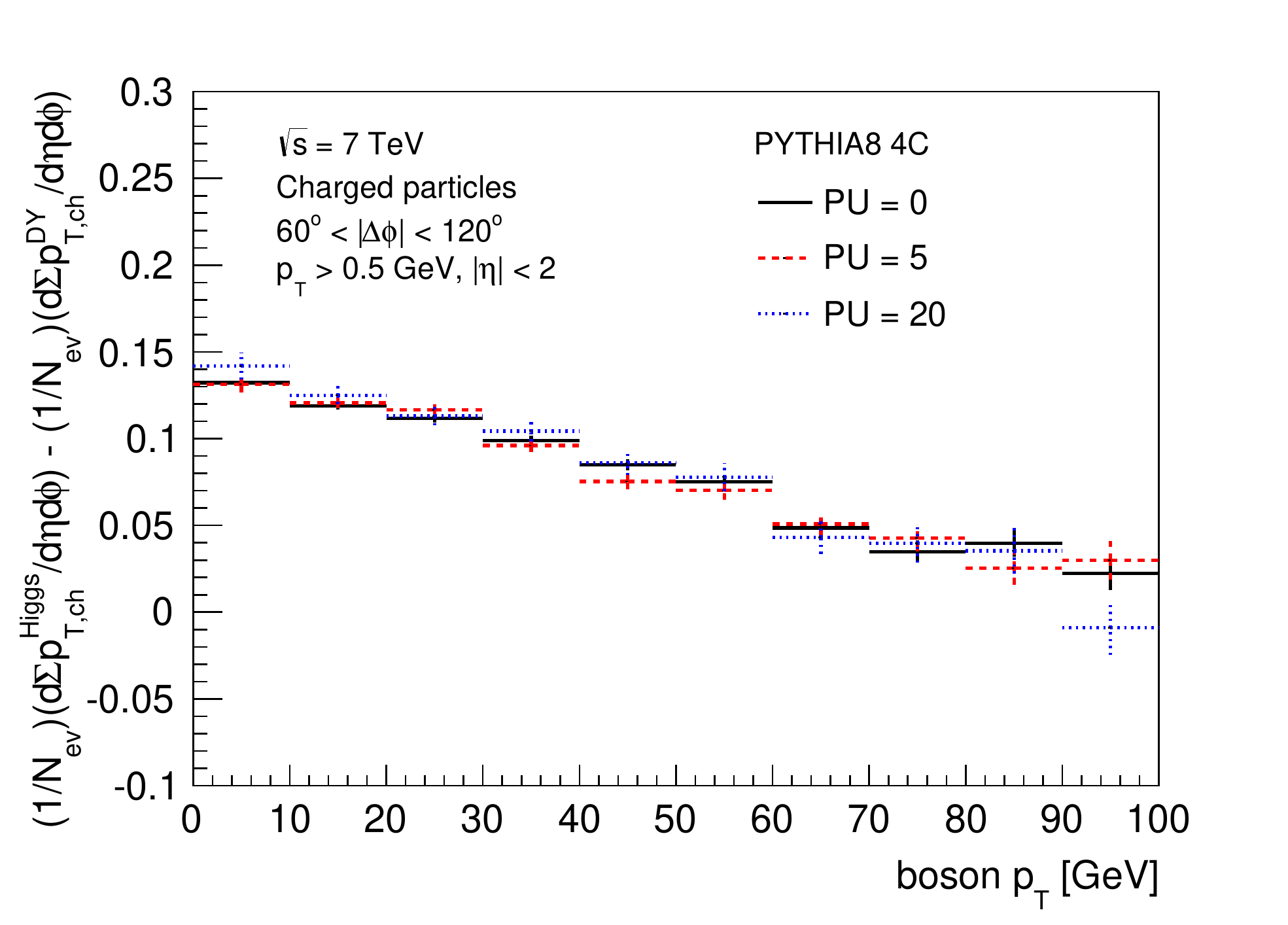}
                \caption{}
                \label{fig:Sub_ChPartPtSum_vs_Bpt}
        \end{subfigure}
        \caption{The subtracted average charged particle multiplicity (a), and average scalar sum of transverse momenta ($\Sigma p_{\text{T}}$) (b) in the transverse region of the azimuthal plane versus transverse momentum $p_{\text{T}}$ of the produced boson. For PU = 0 (solid black), PU = 5 (dashed red) and PU = 20 (dotted blue) scenarios. }
        \label{fig:SubUE}
\end{figure}

A very interesting and widely used observable to study QCD in proton-proton collisions is the underlying event activity \cite{bib:field}. Common measurements \cite{bib:field,bib:CMSUE} look at the charged particle multiplicity and scalar $p_{\text{T}}$ sum in the so called \emph{transverse} region, which is defined as $60^{\text{o}} < \left|\Delta\phi\right| < 120^{\text{o}}$, with $\Delta\phi$ the difference in azimuthal angle between the charged particle and the leading (high $p_{\text{T}}$) object that defines the hard scale of the event. This object can be either the leading track, or jet in the event, but in this study we look at the underlying event activity in Higgs and Drell-Yan production. With the produced boson in the \emph{towards} region ($\left|\Delta\phi\right| < 60^{\text{o}}$), and the recoil jet in the \emph{away} region ($\left|\Delta\phi\right| > 120^{\text{o}}$), the transverse region is only sensitive to the underlying event. Furthermore, when using Higgs and Drell-Yan processes where only the leptonic decay channel of the bosons is considered, one has access to a clean final state, with no contributions from final state radiation (FSR), resulting in a sensitivity to only initial state radiation (ISR) and multiple parton interactions (MPI). Just as with the $p_{\text{T}}$ spectra we can investigate whether we can still measure the underlying event in high PU environments by comparing Higgs and Drell-Yan observables. Obviously, when including additional proton-proton collisions to the event, the number of charged particles (and the scalar $p_{\text{T}}$ sum) will increase drastically and directly scale to the number of PU events. This is illustrated in figures \ref{fig:DefCompPU_ChPartmulti} and \ref{fig:DefCompPU_ChPartmulti_vs_Bpt}, which show the charged particle ($p_{\text{T}} > 0.5$~GeV, $\left|\eta\right| < 2$) multiplicity and average multiplicity as a function of the boson $p_{\text{T}}$ respectively, both with and without additional PU activity. However, we can show that when we subtract the underlying event activity in the Drell-Yan process, from the underlying event in the Higgs process, i.e.:
\begin{equation}
\frac{dn}{dp_{\text{T}}}\left(\text{H} - \text{DY}\right) = \frac{dn}{dp_{\text{T}}^{\text{H}}} + \frac{dn}{dp_{\text{T}}^{\text{PU}}} - \left(\frac{dn}{dp_{\text{T}}^{\text{DY}}} + \frac{dn}{dp_{\text{T}}^{\text{PU}}}\right),  \label{eq:UE}
\end{equation}
the PU contribution completely cancels out. This is confirmed in figures \ref{fig:Sub_ChPartmulti_vs_Bpt} (average charged particle multiplicity versus $p_{\text{T}}^{\text{boson}}$) and \ref{fig:Sub_ChPartPtSum_vs_Bpt} (average charged particle scalar $\Sigma p_{\text{T}}$ versus $p_{\text{T}}^{\text{boson}}$) that show the result of the subtraction. For both additional PU activity of 5 and 20 the subtracted underlying event stays stable. This implies that even in high PU environments one is able to measure small-$p_{\text{T}}$ QCD physics. It is possible to subtract the PU contribution and study the difference in underlying event activity between Higgs and Drell-Yan production, which is directly sensitive to gluon versus quark induced ISR.

\section{Summary}
The Higgs $gg \rightarrow H$ production process provides new perspectives for interesting and novel QCD measurements that allow us to directly probe gluon physics. This is possible due to the electroweak current, which, in the heavy top limit, directly couples to gluons. In addition, the colour singlet state and leptonic decay channels allow us to study events in which no complications from final state effects occur. In this study we presented a novel method that compares Higgs and Drell-Yan production in the same invariant mass and rapidity range to perform a direct measurement of gluon versus quark induced processes. We presented generator studies, performed with the Pythia8 Monte Carlo event generator. First we investigated the inclusive, boson + 1 jet, and boson + 2 jet $p_{\text{T}}$ spectra, and their according Higgs/DY ratios. The inclusive spectrum is found to be sensitive to soft multi-gluon emissions, and is stable in high PU environments. The boson + jet event topologies however suffer from PU effects, that originate from the jet mismatching between signal and PU events. In addition we also studied the underlying event activity in Higgs and Drell-Yan processes by looking at the charged particle multiplicity and scalar $\Sigma p_{\text{T}}$ in the transverse region. We constructed the subtracted underlying event observable (see eq. \ref{eq:UE}) and evaluated its stability in high PU environments. Comparing PU= 0, 5, and 20 scenarios we can conclude that the Higgs - DY subtracted underlying event is indeed stable in high PU environments. Thus, by comparing Higgs to Drell-Yan production we can subtract the PU contributions, and directly measure the difference in gluon versus quark induced initial state radiation effects. Such that one can still access (small-$p_{\text{T}}$) QCD physics in high pile-up environments. 

\section{Acknowledgements}
We are grateful to the organisers, convenors and participants of the DIS conference for this opportunity to present and discuss our results in the QCD and Hadronic Final States working group.

\end{document}